\documentclass[english,aps,superscriptaddress]{revtex4}
\usepackage[T1]{fontenc}
\usepackage[latin9]{inputenc}
\usepackage{mathrsfs}
\usepackage{amsmath}
\usepackage{graphicx}
\usepackage{esint}

\makeatletter

\providecommand{\tabularnewline}{\\}

\@ifundefined{textcolor}{}
{%
 \definecolor{BLACK}{gray}{0}
 \definecolor{WHITE}{gray}{1}
 \definecolor{RED}{rgb}{1,0,0}
 \definecolor{GREEN}{rgb}{0,1,0}
 \definecolor{BLUE}{rgb}{0,0,1}
 \definecolor{CYAN}{cmyk}{1,0,0,0}
 \definecolor{MAGENTA}{cmyk}{0,1,0,0}
 \definecolor{YELLOW}{cmyk}{0,0,1,0}
}

\makeatother

\usepackage{babel}
\begin{document}

\title{Alternative methods to measure global polarization of $\Lambda$
hyperons}

\author{Irfan Siddique}

\affiliation{Department of Modern Physics, University of Science and Technology
of China, Hefei, Anhui 230026, China}

\author{Zuo-tang Liang}

\affiliation{School of Physics \& Key Laboratory of Particle Physics and Particle
Irradiation (MOE)}

\author{Michael Annan Lisa}

\affiliation{Physics Department, The Ohio State University, Columbus, Ohio 43210,
USA}

\author{Qun Wang}

\affiliation{Department of Modern Physics, University of Science and Technology
of China, Hefei, Anhui 230026, China}

\author{Zhang-bu Xu}

\affiliation{Physics Department, Brookhaven National Laboratory, Upton, New York
11973-5000, USA}

\affiliation{School of Physics \& Key Laboratory of Particle Physics and Particle
Irradiation (MOE)}
\begin{abstract}
We propose alternative methods to measure the global polarization
of $\Lambda$ hyperons. These methods involve event averages of proton's
and $\Lambda$'s momenta in the lab frame. We carry out simulations
using these methods and show that all of them work equivalently well
in obtaining the global polarization of $\Lambda$ hyperons. 
\end{abstract}
\maketitle

\section{Introduction}

It is well-known that rotation and polarization are inherently correlated:
the rotation of an uncharged object can lead to spontaneous magnetization
and polarization, and vice versa \cite{Barnett:1935,dehaas:1915}.
We expect that the same phenomena exist in heavy ion collisions. It
is straightforward to estimate that huge global angular momenta are
generated in non-central heavy ion collisions at high energies \cite{Liang:2004ph,Liang:2004xn,Voloshin:2004ha,Betz2007,Becattini:2007sr,Gao2008}.
How such huge global angular momenta are converted to the particle
polarization in the hot and dense matter and how to measure the global
polarization are two core questions to be answered. To address the
first question, there are some theoretical models in the market, e.g.,
the microscopic spin-orbital coupling model \cite{Liang:2004ph,Liang:2004xn,Gao2008,Chen:2008wh},
the statistical-hydro model \cite{Becattini:2009wh,Becattini:2012tc,Becattini:2013fla,Becattini:2015nva}
and the kinetic model with Wigner functions \cite{Gao:2012ix,Chen:2012ca,Fang:2016vpj,Fang:2016uds},
see Ref. \cite{Wang:2017jpl} for a recent review. For the second
question, one can use the weak decay property of $\Lambda$ hyperons
to measure the global polarization \cite{Liang:2004ph,Liang:2004xn}:
the parity-breaking weak decay of $\Lambda$ into a proton and a pion
is self-analysing since the daughter proton is emitted preferentially
along $\Lambda$'s spin in $\Lambda$'s rest frame \cite{Overseth:1967zz,Voloshin:2004ha}.
The global polarization of a vector meson can be measured through
the angular distribution of its decay products which is related to
some elements of its spin density matrix \cite{Liang:2004xn}. 

Recently the global polarization of $\Lambda$ and $\bar{\Lambda}$
has been measured at collisional energies below 62.4 GeV \cite{STAR:2017ckg,Abelev:2007zk}.
The average values of the global polarization for $\Lambda$ and $\bar{\Lambda}$
are $\mathscr{P}_{\Lambda}=(1.08\pm0.15)\%$ and $\mathscr{P}_{\bar{\Lambda}}=(1.38\pm0.30)\%$.
The polarization of $\bar{\Lambda}$ is a little larger than that
of $\Lambda$ which is thought to be caused by a negative (positive)
magnetic moment of $\Lambda$($\bar{\Lambda}$) in magnetic fields.
But such a difference is negligible within the error bars and magnetic
fields extracted from the data are consistent to zero. The global
polarization of $\Lambda$ and $\bar{\Lambda}$ decreases with collisional
energies. This is due to that the Bjorken scaling works better at
higher energies than lower energies. From the data one can estimate
the local vorticity: $\omega=(9\pm1)\times10^{21}\:\mathrm{s}^{-1}$,
implying that the matter created in ultra-relativistic heavy ion collisions
is the most vortical fluid that ever exists in nature. The vorticity
field of the quark gluon plasma has been studied by many authors in
a variety of methods including hydrodynamical models \cite{Csernai:2013bqa,Csernai:2014ywa,Pang:2016igs}
and transport models \cite{Jiang:2016woz,Deng:2016gyh}. The global
polarization of $\Lambda$ and $\bar{\Lambda}$ has also been calculated
by hydrodynamical models \cite{Karpenko:2016jyx,Xie:2017upb}, the
transport model \cite{Li:2017slc} the chiral kinetic model \cite{Sun:2017xhx}. 

The method used in the STAR measurement is through the event average
of $\sin\left(\phi_{\mathrm{p}}^{*}-\psi_{\mathrm{RP}}\right)$, where
$\phi_{\mathrm{p}}^{*}$ and $\psi_{\mathrm{RP}}$ are the azimuthal
angle of the proton momemtum in $\Lambda$'s rest frame and that of
the reaction plane respectively \cite{STAR:2017ckg,Abelev:2007zk}.
The orientation of the reaction plane cannot be directly measured
but through that of the event plane determined from the direct flow.
Therefore a reaction plane resolution factor was introduced to account
for the finite resolution of the reaction plane by the detector \cite{STAR:2017ckg,Abelev:2007zk}. 

In this paper, we propose alternative methods to measure the global
polarization of $\Lambda$ and $\bar{\Lambda}$ hyperons based on
Lorentz transformation. The advantages of these methods are that all
event averages are taken over momenta in the lab frame instead of
$\Lambda$'s rest frame. We compare these methods by simulations and
show that all of them work equivalently well in obtaining the global
polarization of $\Lambda$ hyperons.

\section{Hyperon's weak decay and polarization}

The polarization of the $\Lambda$ (and $\bar{\Lambda}$) hyperons
can be measured by its parity-breaking weak decay $\Lambda\rightarrow\mathrm{p}+\pi^{-}$.
The daughter protons are emitted preferentially along the $\Lambda$'s
polarization in $\Lambda$'s rest frame. The angular distribution
of the daughter proton reads
\begin{equation}
\frac{dN}{d\Omega^{*}}=\frac{1}{4\pi}\left(1+\alpha_{H}\mathscr{P}_{\Lambda}\frac{\mathbf{n}^{*}\cdot\mathbf{p}^{*}}{|\mathbf{p}^{*}|}\right),\label{eq:dn-doemga}
\end{equation}
where $\alpha_{H}$ is the hyperon decay parameter, $\mathscr{P}_{\Lambda}$
is the Lambda global polarization, $\mathbf{n}^{*}$, $\mathbf{p}^{*}$
and $\Omega^{*}$ are the $\Lambda$'s polarization, the proton's
momentum and its solid angles respectively in the rest frame of the
hyperon which are labeled by the superscript $'*'$. We note that
Eq. (\ref{eq:dn-doemga}) is Lorentz invariant by observing 
\begin{eqnarray}
\mathbf{n}^{*}\cdot\mathbf{p}^{*} & = & -n_{\mu}p^{\mu}=-n\cdot p,\nonumber \\
E_{\mathrm{p}}^{*} & = & \frac{1}{2m_{\Lambda}}(m_{\Lambda}^{2}+m_{\mathrm{p}}^{2}-m_{\pi}^{2}),\nonumber \\
|\mathbf{p}^{*}| & = & \frac{1}{2m_{\Lambda}}\sqrt{[m_{\Lambda}^{2}-(m_{\mathrm{p}}-m_{\pi})^{2}][m_{\Lambda}^{2}-(m_{\mathrm{p}}+m_{\pi})^{2}]},\label{eq:p-star-e-star}
\end{eqnarray}
where $p^{\mu}$ and $p_{\Lambda}^{\mu}$ are the four-momentum of
the proton and the hyperon in any frame respectively, $n^{\mu}$ is
the space-like four-vector of the hyperon's polarization in a general
frame. We now focus on the lab frame and the hyperon's rest frame.
We now use $p^{\mu}$, $p_{\Lambda}^{\mu}$ and $n^{\mu}$ to label
quantities in the lab frame and all quantities with the superscript
$'*'$ are those in the hyperon's rest frame. We have Lorentz transformation
for the $\Lambda$'s polarization, 
\begin{equation}
n^{\mu}=\Lambda_{\;\nu}^{\mu}(-\mathbf{v}_{\Lambda})n^{*\nu},\label{eq:lorentz-tr}
\end{equation}
where $\Lambda_{\;\nu}^{\mu}(-\mathbf{v}_{\Lambda})$ is Lorentz transformation
with $\mathbf{v}_{\Lambda}=\mathbf{p}_{\Lambda}/E_{\Lambda}$. The
$\Lambda$'s polarization in the rest frame $n^{*\nu}$ has the form
$n^{*\mu}=(0,\mathbf{n}^{*})$ where $\mathbf{n}^{*}$ is the three-vector
of the polarization with $|\mathbf{n}^{*}|^{2}<1$. From Eq. (\ref{eq:lorentz-tr})
we have 
\begin{equation}
n^{\mu}=(n_{0},\mathbf{n})=\left(\frac{\mathbf{n}^{*}\cdot\mathbf{p}_{\Lambda}}{m_{\Lambda}},\mathbf{n}^{*}+\frac{(\mathbf{n}^{*}\cdot\mathbf{p}_{\Lambda})\mathbf{p}_{\Lambda}}{m_{\Lambda}(m_{\Lambda}+E_{\Lambda})}\right).\label{eq:pol-rest}
\end{equation}
We can also express $n^{*\mu}$ in terms of $n^{\mu}$, 
\begin{eqnarray}
n^{*\mu} & = & \Lambda_{\;\nu}^{\mu}(\mathbf{v}_{\Lambda})n^{\nu},
\end{eqnarray}
or explicitly, 
\begin{eqnarray}
n^{*\mu}=(0,\mathbf{n}^{*}) & = & \left(0,\mathbf{n}-\frac{\mathbf{p}_{\Lambda}(\mathbf{n}\cdot\mathbf{p}_{\Lambda})}{E_{\Lambda}(E_{\Lambda}+m_{\Lambda})}\right).
\end{eqnarray}

The polarization four-vector of one particle is always orthogonal
to its four-momentum, $n\cdot p_{\Lambda}=n^{0}E_{\Lambda}-\mathbf{n}\cdot\mathbf{p}_{\Lambda}=0$,
so we can express $n^{0}$ in term of $\mathbf{n}$, $n^{0}=\mathbf{n}\cdot\mathbf{v}_{\Lambda}$.
One can verify that $n^{\mu}$ in Eq. (\ref{eq:pol-rest}) does satisfy
$n^{0}=\mathbf{n}\cdot\mathbf{v}_{\Lambda}$. From $(n^{0})^{2}-|\mathbf{n}|^{2}=-|\mathbf{n}^{*}|^{2}$
and $n^{0}=\mathbf{n}\cdot\mathbf{v}_{\Lambda}$, we can solve $|\mathbf{n}|^{2}$
as 
\begin{equation}
|\mathbf{n}|^{2}=\frac{|\mathbf{n}^{*}|^{2}}{1-|\mathbf{v}_{\Lambda}|^{2}(\hat{\mathbf{n}}\cdot\hat{\mathbf{v}}_{\Lambda})^{2}}.
\end{equation}
We see that when $|\mathbf{v}_{\Lambda}|^{2}(\hat{\mathbf{n}}\cdot\hat{\mathbf{v}}_{\Lambda})^{2}\rightarrow1$,
$|\mathbf{n}|^{2}\rightarrow\infty$, so $|\mathbf{n}|^{2}$ is not
bound. In case of transverse polarization, i.e. $\hat{\mathbf{n}}\cdot\hat{\mathbf{v}}_{\Lambda}=0$,
we have $|\mathbf{n}|^{2}=|\mathbf{n}^{*}|^{2}<1$. 

In the lab frame, a 3-dimensional vector (e.g. impact parameter, global
angular momentum, beam) can be written as $\mathbf{a}=\mathbf{a}_{x}\mathbf{e}_{x}+\mathbf{a}_{y}\mathbf{e}_{y}+\mathbf{a}_{z}\mathbf{e}_{z}$
with $(\mathbf{e}_{x},\mathbf{e}_{y},\mathbf{e}_{z})$ being three
basis directions.

\section{Previous method to measure hyperon's polarization}

In this section we introduce the method used in STAR's measurement
of the $\Lambda$'s polarization \cite{Abelev:2007zk}. From Eq. (\ref{eq:n-star-p-star})
we can determine the $\Lambda$'s polarization in the rest frame by
taking the event average over the direction of the proton momentum
$\hat{\mathbf{p}}^{*}$. Then we make projection onto the direction
of the global angular momentum $\mathbf{e}_{L}$, 
\begin{eqnarray}
\mathscr{P}_{\Lambda} & = & \frac{3}{\alpha_{H}}\left\langle \hat{\mathbf{p}}^{*}\cdot\mathbf{e}_{L}\right\rangle _{\mathrm{ev}}\nonumber \\
 & = & \frac{3}{\alpha_{H}}\left\langle \cos\theta^{*}\right\rangle _{\mathrm{ev}}\label{eq:p-lambda}
\end{eqnarray}
where $\theta^{*}$ is the angle in $\Lambda$'s rest frame between
the proton momentum and the global angular momentum corresponding
to the reaction plane. We have the following relation 
\begin{equation}
\cos\theta^{*}=\sin\theta_{\mathrm{p}}^{*}\sin\left(\phi_{\mathrm{p}}^{*}-\psi_{\mathrm{RP}}\right),
\end{equation}
where $\theta_{\mathrm{p}}^{*}$ and $\phi_{\mathrm{p}}^{*}$ are
the polar and azimuthal angle of $\hat{\mathbf{p}}^{*}$ respectively,
and $\psi_{\mathrm{RP}}$ is the azimuthal angle of the reaction plane.
Then we can integrate over $\theta_{\mathrm{p}}^{*}$ of Eq. (\ref{eq:dn-doemga})
to obtain 
\begin{eqnarray}
\frac{dN}{d\phi_{p}^{*}} & = & \int_{0}^{\pi}d\theta_{\mathrm{p}}^{*}\sin\theta_{\mathrm{p}}^{*}\frac{dN}{d\Omega^{*}}=\frac{1}{2\pi}+\frac{1}{8}\alpha_{H}\mathscr{P}_{\Lambda}\sin\left(\phi_{\mathrm{p}}^{*}-\psi_{\mathrm{RP}}\right),
\end{eqnarray}
which gives the polarization in terms of the azimuthal angle of the
daughter proton, 
\begin{eqnarray}
\mathscr{P}_{\Lambda} & = & \frac{8}{\pi\alpha_{H}}\left\langle \sin\left(\phi_{\mathrm{p}}^{*}-\psi_{\mathrm{RP}}\right)\right\rangle _{\mathrm{ev}},\label{eq:spin}
\end{eqnarray}
with 
\begin{equation}
\left\langle \sin\left(\phi_{\mathrm{p}}^{*}-\psi_{\mathrm{RP}}\right)\right\rangle _{\mathrm{ev}}=\int_{0}^{2\pi}d\phi_{p}^{*}\frac{dN}{d\phi_{p}^{*}}\sin\left(\phi_{\mathrm{p}}^{*}-\psi_{\mathrm{RP}}\right).
\end{equation}

In the STAR experiment, the azimuthal angle of the reaction plane
cannot be directly measured but through the measurement of the event
plane by direct flows. This introduces a reaction plane resolution
factor in the denominator in Eq. (\ref{eq:spin}), $R_{\mathrm{EP}}^{(1)}=\left\langle \cos\left(\psi_{\mathrm{RP}}-\psi_{\mathrm{EP}}^{(1)}\right)\right\rangle _{\mathrm{ev}}$,
where $\psi_{\mathrm{EP}}^{(1)}$ is the azimuthal angle of the event
plane determined by the direct flows.

\section{Alternative methods}

In this section, we introduce alternative methods to measure the $\Lambda$'s
polarization. The advantage of these methods is that the polarization
can be measured through the proton's momentum in the lab frame. 

\begin{figure}
\caption{\label{fig:coordinate}In the coordinate system ($x,y,z$), the beam
direction is along the negative z-direction, the impact parameter
vector is in the x-direction, and the orbital angular momentum is
in the y-direction. The direction of the proton momentum can be described
by the polar angle $\theta_{\mathrm{p}}$ and the azimuthal angle
$\phi_{\mathrm{p}}$. The coordinate system ($x^{\prime},y^{\prime},z^{\prime}$)
is used in experiment. The $z^{\prime}$-axis is just the z-axis.
The azimuthal angle of the impact parameter vector in the ($x^{\prime},y^{\prime},z^{\prime}$)
system is $\psi_{\mathrm{RP}}$. }

\includegraphics[scale=0.4]{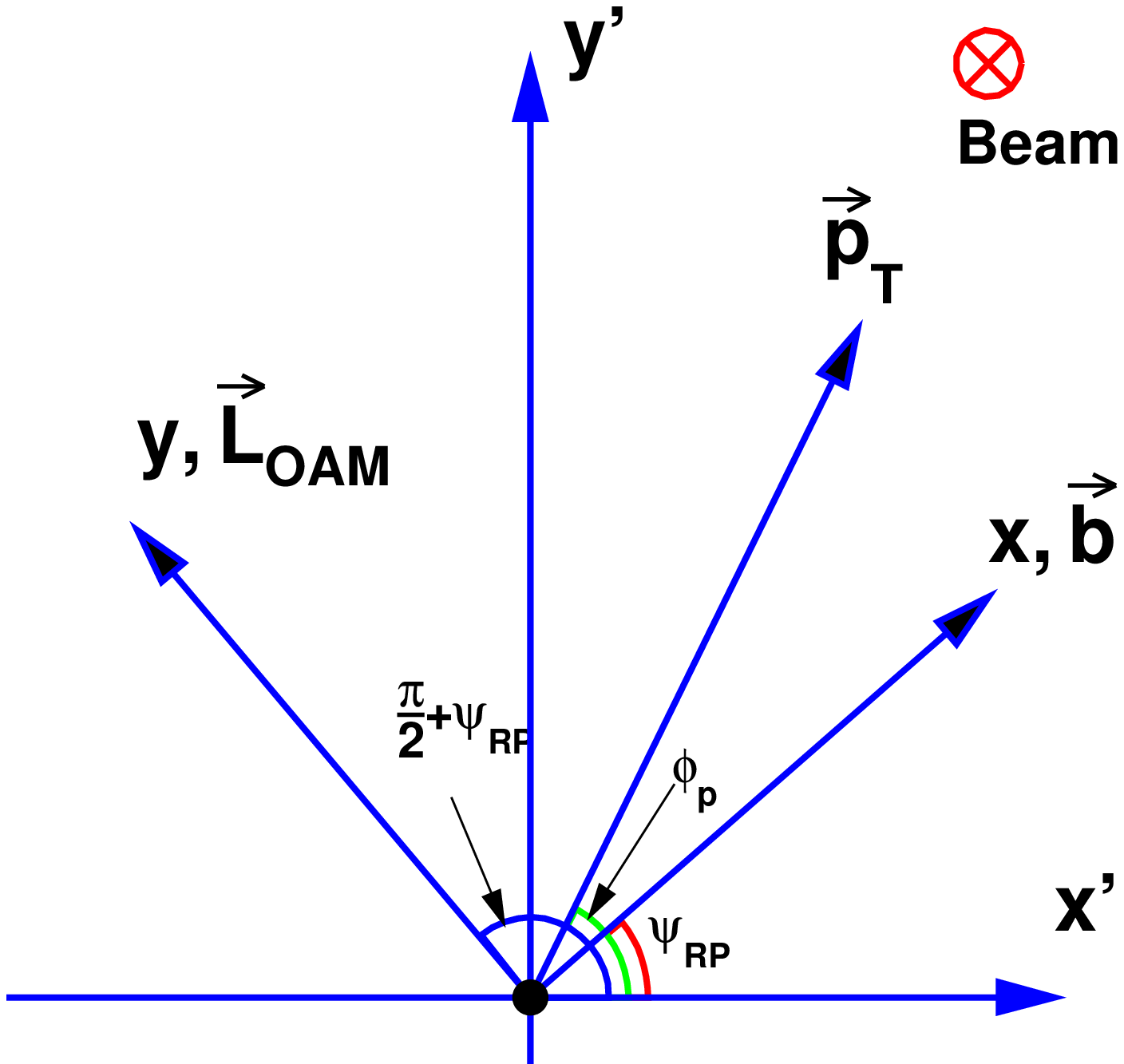}\includegraphics[scale=0.4]{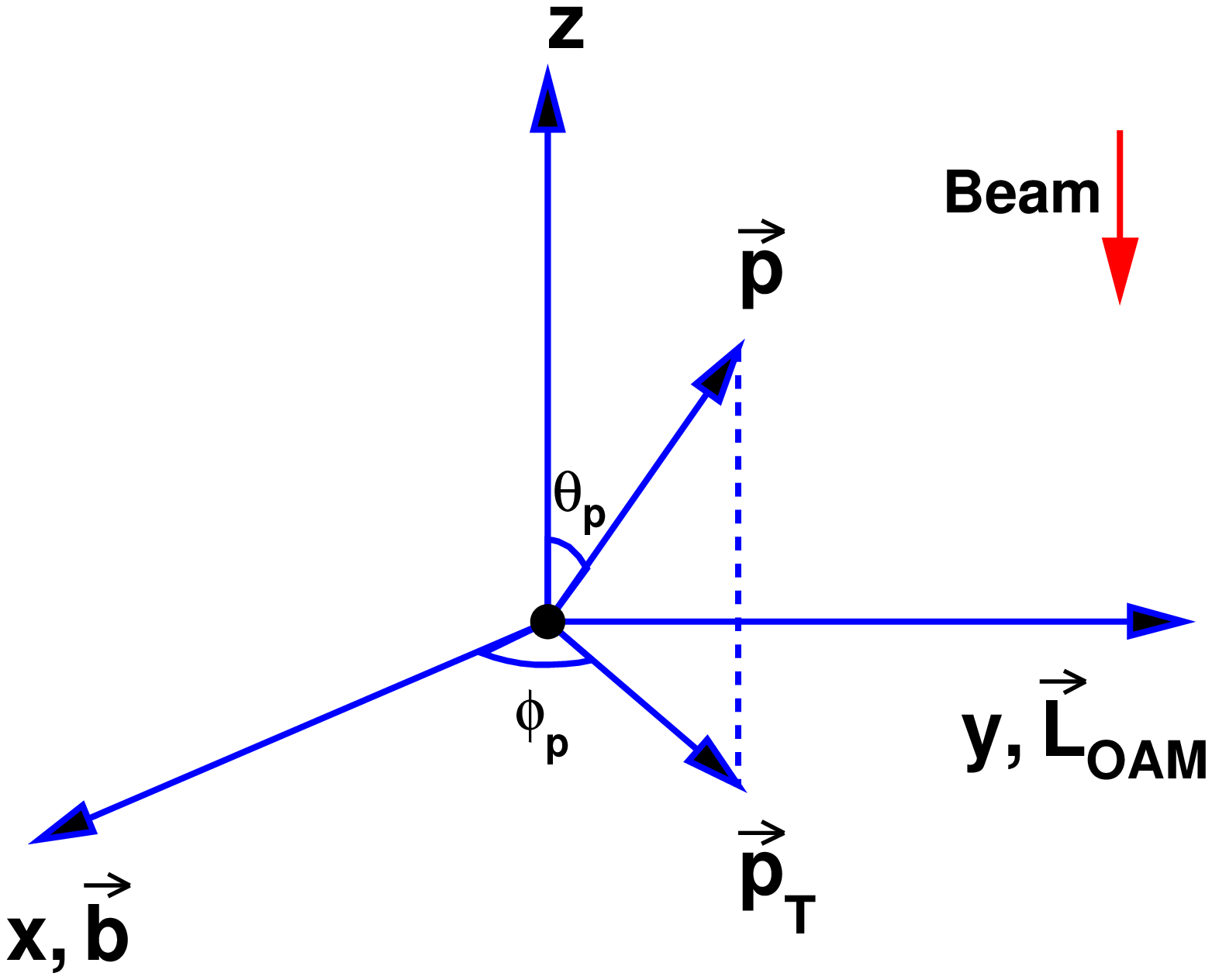}
\end{figure}

We start with the formula for $\Lambda$'s polarization vector in
its rest frame, 
\begin{equation}
\overrightarrow{\mathscr{P}}_{\Lambda}=\frac{3}{\alpha_{H}}\left\langle \hat{\mathbf{p}}^{*}\right\rangle _{\mathrm{ev}}.\label{eq:n-star-p-star}
\end{equation}
We can project the above onto the direction of the global polarization
which we assume to be along the y-axis, see Fig. (\ref{fig:coordinate}). 

Now we try to evaluate $\left\langle \hat{\mathbf{p}}^{*}\right\rangle _{\mathrm{ev}}$.
To this end, we use following Lorentz transformation for the proton's
momentum, 
\begin{eqnarray}
\mathbf{p} & = & \mathbf{p}^{*}+\frac{\mathbf{p}_{\Lambda}(\mathbf{p}^{*}\cdot\mathbf{p}_{\Lambda})}{m_{\Lambda}\left(E_{\Lambda}+m_{\Lambda}\right)}+\frac{E_{\mathrm{p}}^{*}}{m_{\Lambda}}\mathbf{p}_{\Lambda},\label{eq:p-p-star}
\end{eqnarray}
where $E_{\mathrm{p}}^{*}$ is determined by the masses of the proton,
pion and $\Lambda$ as in Eq. (\ref{eq:p-star-e-star}). Now we take
the event average of $\left\langle \mathbf{p}\right\rangle _{\mathrm{ev}}$,
\begin{eqnarray}
\left\langle \mathbf{p}\right\rangle _{\mathrm{ev}} & = & \left\langle \mathbf{p}^{*}\right\rangle _{\mathrm{ev}}+\left\langle \frac{\mathbf{p}_{\Lambda}(\mathbf{p}^{*}\cdot\mathbf{p}_{\Lambda})}{m_{\Lambda}\left(E_{\Lambda}+m_{\Lambda}\right)}\right\rangle _{\mathrm{ev}},\label{eq:p-ev}
\end{eqnarray}
where we have used $\left\langle \mathbf{p}_{\Lambda}\right\rangle _{\mathrm{ev}}=0$. 

In order to evaluate the second event average in the right-hand-side
of Eq. (\ref{eq:p-ev}), we make two assumptions: (1) $\mathbf{p}_{\Lambda}$
and $\mathbf{p}^{*}$ are statistically independent, so we have $\left\langle \mathbf{p}_{\Lambda}(\mathbf{p}_{\Lambda}\cdot\mathbf{p}^{*})\right\rangle _{\mathrm{ev}}\approx\mathbf{e}_{i}\left\langle \mathbf{p}_{\Lambda}^{i}\mathbf{p}_{\Lambda}^{j}\right\rangle _{\mathrm{ev}}\left\langle \mathbf{p}_{j}^{*}\right\rangle _{\mathrm{ev}}$,
where $\mathbf{p}_{\Lambda}=\mathbf{e}_{i}\mathbf{p}_{\Lambda}^{i}$
with $i=x,y,z$; (2) $\left\langle \mathbf{p}_{\Lambda}^{i}\mathbf{p}_{\Lambda}^{j}\right\rangle _{\mathrm{ev}}=\left\langle |\mathbf{p}_{\Lambda}^{i}|^{2}\right\rangle _{\mathrm{ev}}\delta_{ij}$.
Then Eq. (\ref{eq:p-ev}) becomes 
\begin{eqnarray}
\left\langle \hat{\mathbf{p}}_{x}^{*}\right\rangle _{\mathrm{ev}} & \approx & \frac{1}{|\mathbf{p}^{*}|}\left(1+\left\langle \frac{|\mathbf{p}_{\Lambda}^{x}|^{2}}{\left(E_{\Lambda}+m_{\Lambda}\right)m_{\Lambda}}\right\rangle _{\mathrm{ev}}\right)^{-1}\left\langle \mathbf{p}_{x}\right\rangle _{\mathrm{ev}},\nonumber \\
\left\langle \hat{\mathbf{p}}_{y}^{*}\right\rangle _{\mathrm{ev}} & \approx & \frac{1}{|\mathbf{p}^{*}|}\left(1+\left\langle \frac{|\mathbf{p}_{\Lambda}^{y}|^{2}}{\left(E_{\Lambda}+m_{\Lambda}\right)m_{\Lambda}}\right\rangle _{\mathrm{ev}}\right)^{-1}\left\langle \mathbf{p}_{y}\right\rangle _{\mathrm{ev}},\nonumber \\
\left\langle \hat{\mathbf{p}}_{z}^{*}\right\rangle _{\mathrm{ev}} & \approx & \frac{1}{|\mathbf{p}^{*}|}\left(1+\left\langle \frac{|\mathbf{p}_{\Lambda}^{z}|^{2}}{\left(E_{\Lambda}+m_{\Lambda}\right)m_{\Lambda}}\right\rangle _{\mathrm{ev}}\right)^{-1}\left\langle \mathbf{p}_{z}\right\rangle _{\mathrm{ev}}.\label{eq:p-star}
\end{eqnarray}
Now we choose the new coordinate system as in Fig. (\ref{fig:coordinate}):
the impact parameter vector is along the x-axis, the global orbital
momentum is along the y-axis and the beam direction is along the negative
z-axis. The old coordinate system is the one that is used in the experiment:
the beam direction is along the negative z-axis, the impact parameter
vector (reaction plane) has azimuthal angle $\psi_{\mathrm{RP}}$
relative to the x-axis. In the new coordinate system, we have $\mathbf{p}_{\Lambda,\mathrm{p}}^{x}=|\mathbf{p}_{\Lambda,\mathrm{p}}^{T}|\cos\left(\phi_{\Lambda,\mathrm{p}}-\psi_{\mathrm{RP}}\right)$
and $\mathbf{p}_{\Lambda,\mathrm{p}}^{y}=|\mathbf{p}_{\Lambda,\mathrm{p}}^{T}|\sin\left(\phi_{\Lambda,\mathrm{p}}-\psi_{\mathrm{RP}}\right)$
with $\phi_{\Lambda,\mathrm{p}}$ being the azimuthal angle of the
$\Lambda$ hyperon and proton respectively. 

We can further simplifiy Eq. (\ref{eq:p-star}) by using the elliptic
flow coefficients. The distribution of $\mathbf{p}_{\Lambda}$ is
not isotropic but satisfies 
\begin{eqnarray}
\left\langle |\mathbf{p}_{\Lambda}^{x}|^{2}\right\rangle _{\mathrm{ev}} & \approx & \left\langle |\mathbf{p}_{\Lambda}^{T}|^{2}\right\rangle _{\mathrm{ev}}\left\langle \cos^{2}\left(\phi_{\Lambda}-\psi_{\mathrm{RP}}\right)\right\rangle _{\mathrm{ev}}\approx\left\langle |\mathbf{p}_{\Lambda}^{T}|^{2}\right\rangle _{\mathrm{ev}}\frac{1}{2}\left(1+v_{2}^{\Lambda}\right),\nonumber \\
\left\langle |\mathbf{p}_{\Lambda}^{y}|^{2}\right\rangle _{\mathrm{ev}} & \approx & \left\langle |\mathbf{p}_{\Lambda}^{T}|^{2}\right\rangle _{\mathrm{ev}}\left\langle \sin^{2}\left(\phi_{\Lambda}-\psi_{\mathrm{RP}}\right)\right\rangle _{\mathrm{ev}}\approx\left\langle |\mathbf{p}_{\Lambda}^{T}|^{2}\right\rangle _{\mathrm{ev}}\frac{1}{2}\left(1-v_{2}^{\Lambda}\right),
\end{eqnarray}
where $v_{2}^{\Lambda}$ is the elliptic flow of the $\Lambda$ hyperon.
Since the global angular momentum is along the y-axis, we have $\left\langle \mathbf{p}_{x}\right\rangle _{\mathrm{ev}}=\left\langle \mathbf{p}_{z}\right\rangle _{\mathrm{ev}}=0$,
so only non-vanishing component is 
\begin{eqnarray}
\left\langle \hat{\mathbf{p}}_{y}^{*}\right\rangle _{\mathrm{ev}} & \approx & \frac{1}{|\mathbf{p}^{*}|}\left(1+\left\langle \frac{|\mathbf{p}_{\Lambda}^{T}|^{2}\sin^{2}\left(\phi_{\Lambda}-\psi_{\mathrm{RP}}\right)}{\left(E_{\Lambda}+m_{\Lambda}\right)m_{\Lambda}}\right\rangle _{\mathrm{ev}}\right)^{-1}\left\langle |\mathbf{p}_{T}|\sin\left(\phi_{\mathrm{p}}-\psi_{\mathrm{RP}}\right)\right\rangle _{\mathrm{ev}}\nonumber \\
 & \approx & \frac{1}{|\mathbf{p}^{*}|}\left[1+\frac{1}{2}\left(1-v_{2}^{\Lambda}\right)\left\langle \frac{|\mathbf{p}_{\Lambda}^{T}|^{2}}{\left(E_{\Lambda}+m_{\Lambda}\right)m_{\Lambda}}\right\rangle _{\mathrm{ev}}\right]^{-1}\left\langle |\mathbf{p}_{T}|\sin\left(\phi_{\mathrm{p}}-\psi_{\mathrm{RP}}\right)\right\rangle _{\mathrm{ev}}\label{eq:angular-av}
\end{eqnarray}

In the central rapidity region we have $|\mathbf{p}_{\Lambda}^{z}|\ll|\mathbf{p}_{\Lambda}^{T}|$
and then $|\mathbf{p}_{\Lambda}^{T}|\approx|\mathbf{p}_{\Lambda}|$,
Eq. (\ref{eq:angular-av}) becomes 
\begin{eqnarray}
\left\langle \hat{\mathbf{p}}_{y}^{*}\right\rangle _{\mathrm{ev}} & \approx & \frac{1}{|\mathbf{p}^{*}|}\left[1+\frac{1}{2}(1-v_{2}^{\Lambda})\left(\left\langle \gamma_{\Lambda}\right\rangle _{\mathrm{ev}}-1\right)\right]^{-1}\left\langle |\mathbf{p}_{T}|\sin\left(\phi_{\mathrm{p}}-\psi_{\mathrm{RP}}\right)\right\rangle _{\mathrm{ev}}\label{eq:angular-av-1}
\end{eqnarray}
In non-relativistic limit when $\gamma_{\Lambda}\approx1$ or $|\mathbf{v}_{\Lambda}|\approx0$,
we obtain 
\begin{eqnarray}
\left\langle \hat{\mathbf{p}}_{y}^{*}\right\rangle _{\mathrm{ev}} & \approx & \frac{1}{|\mathbf{p}^{*}|}\left\langle |\mathbf{p}_{T}|\sin\left(\phi_{\mathrm{p}}-\psi_{\mathrm{RP}}\right)\right\rangle _{\mathrm{ev}}\label{eq:angular-av-2}
\end{eqnarray}
The difference from the previous method is that we now take event
average over the proton's momenta in the lab frame. 

Another method is to use the Lorentz transformation for the energy
associated with Eq. (\ref{eq:p-p-star}) 
\begin{equation}
E_{\mathrm{p}}=\gamma_{\Lambda}E_{\mathrm{p}}^{*}+\frac{\mathbf{p}^{*}\cdot\mathbf{p}_{\Lambda}}{m_{\Lambda}}
\end{equation}
to replace $(\mathbf{p}^{*}\cdot\mathbf{p}_{\Lambda})/m_{\Lambda}$
with $E_{\mathrm{p}}-\gamma_{\Lambda}E_{\mathrm{p}}^{*}$ in Eq. (\ref{eq:p-p-star}).
Then Eq. (\ref{eq:p-p-star}) becomes 
\begin{eqnarray}
\mathbf{p} & = & \mathbf{p}^{*}+(E_{\mathrm{p}}-\gamma_{\Lambda}E_{\mathrm{p}}^{*})\frac{\mathbf{p}_{\Lambda}}{E_{\Lambda}+m_{\Lambda}}+\frac{E_{\mathrm{p}}^{*}}{m_{\Lambda}}\mathbf{p}_{\Lambda}\nonumber \\
 & = & \mathbf{p}^{*}+\frac{E_{\mathrm{p}}}{E_{\Lambda}+m_{\Lambda}}\mathbf{p}_{\Lambda}+\frac{E_{\mathrm{p}}^{*}}{E_{\Lambda}+m_{\Lambda}}\mathbf{p}_{\Lambda}.
\end{eqnarray}
When taking the event average, using $\left\langle \mathbf{p}_{\Lambda}/(E_{\Lambda}+m_{\Lambda})\right\rangle \approx0$,
we obtain 
\begin{eqnarray}
\left\langle \mathbf{p}^{*}\right\rangle _{\mathrm{ev}} & = & \left\langle \mathbf{p}\right\rangle _{\mathrm{ev}}-\left\langle \frac{E_{\Lambda}}{E_{\Lambda}+m_{\Lambda}}E_{\mathrm{p}}\mathbf{v}_{\Lambda}\right\rangle _{\mathrm{ev}}\nonumber \\
 & = & m_{\mathrm{p}}\left\langle \gamma_{\mathrm{p}}\left(\mathbf{v}_{\mathrm{p}}-\frac{\gamma_{\Lambda}}{\gamma_{\Lambda}+1}\mathbf{v}_{\Lambda}\right)\right\rangle _{\mathrm{ev}},\label{eq:p-star-1}
\end{eqnarray}
where $\gamma_{\mathrm{p}}$ and $\gamma_{\Lambda}$ are Lorentz contraction
factors for the proton and $\Lambda$ repectively. We see the right-hand-side
of the above equation involves only momenta in the lab frame. We can
project Eq. (\ref{eq:p-star-1}) onto the y-direction (the direction
of the orbital angular momentum) to obtain $\left\langle \mathbf{p}_{y}^{*}\right\rangle _{\mathrm{ev}}$. 

With $\left\langle \hat{\mathbf{p}}_{y}^{*}\right\rangle _{\mathrm{ev}}$
given in one of Eqs. (\ref{eq:angular-av},\ref{eq:angular-av-1},\ref{eq:p-star-1}),
we can obtain the global polarization of $\Lambda$ from through Eq.
(\ref{eq:n-star-p-star}). In the next section we will compare these
methods by simulations.

\section{Simulation results with UrQMD}

\begin{figure}
\caption{\label{fig:rapidity}The dependence of simulation results on rapidity
ranges for the global polarizaton of the $\Lambda$ hyperon. The same
parameters and number of events are used as in Table \ref{tab:1}. }

\includegraphics[scale=0.8]{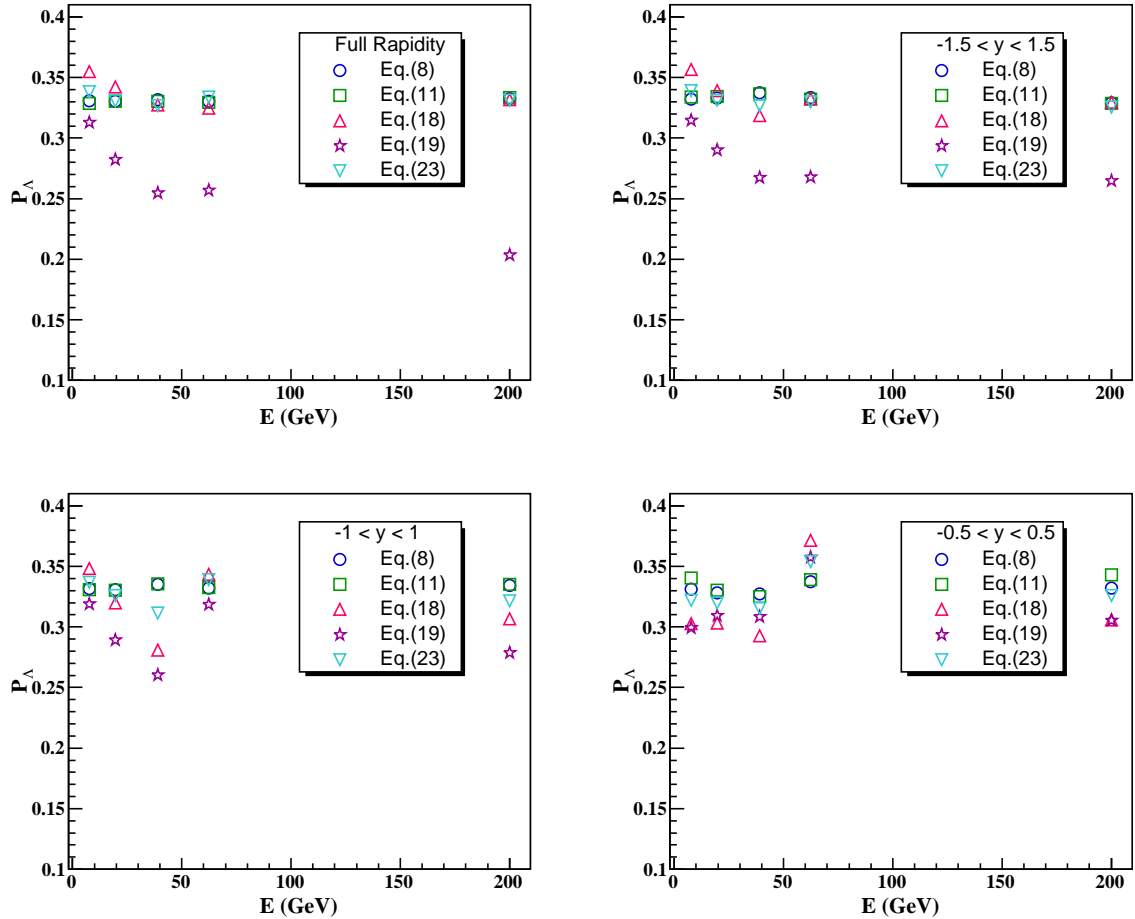}
\end{figure}

The UrQMD model \cite{Bass:1998ca,Petersen:2008dd} has been used
for producing an ensemble of $\Lambda$'s four-momemta $(E_{\Lambda},\mathbf{p}_{\Lambda})$
for Au+Au collisions with the impact parameter 6 fm for collisional
energies listed in Table \ref{tab:1}. In each event there are a few
$\Lambda$ hyperons produced. All these hyperons are collected. Each
hyperon is allowed to decay into a proton and a pion whose angular
distribution in $\Lambda$'s rest frame is given by 
\begin{equation}
\frac{dN}{d\Omega^{*}}=\frac{1}{4\pi}\left(1+\alpha_{H}\mathscr{P}_{\Lambda}\frac{\mathbf{n}^{*}\cdot\mathbf{p}^{*}}{|\mathbf{p}^{*}|}\right),\label{eq:dn-doemga-1}
\end{equation}
where $\mathscr{P}_{\Lambda}$ denotes the $\Lambda$'s polarization.
By taking a specific value of $\mathscr{P}_{\Lambda}$, we can then
sample proton's momenta in $\Lambda$'s rest frames. For each $\varLambda$
hyperon, the proton's momentum in its rest frame is then boosted back
to the lab frame. In this way we create an ensemble of proton momenta
in the lab frame. With the ensemble of momenta for protons and $\Lambda$s,
we can obtain $\left\langle \mathbf{p}_{y}^{*}\right\rangle _{\mathrm{ev}}$.
Here we choose the direction of the global angular momentum along
the y-direction. Finally we obtain $\mathscr{P}_{\Lambda}$ by Eq.
(\ref{eq:n-star-p-star}). The simulation results for the global polarizaton
of $\Lambda$ hyperons using methods corresponding to Eqs. (\ref{eq:p-lambda},\ref{eq:spin},\ref{eq:angular-av},\ref{eq:p-star-1},\ref{eq:angular-av-1})
are shown in Table \ref{tab:1}. We see in the table that all these
methods work equivalently well as the STAR method. Figure \ref{fig:rapidity}
shows the dependence of simulation results on rapidity ranges. We
can see that all these methods work well for the rapidity ranges chosen
except that using Eq. (\ref{eq:angular-av-1}) in the full rapidity
range, $[-1.5,1.5]$ and $[-1,1]$. This is understandable since Eq.
(\ref{eq:angular-av-1}) is only valid in central rapidity. We can
see that the method using Eq. (\ref{eq:angular-av-1}) does work well
in the central rapidity range $[-0.5.0.5]$. 

\begin{table}
\caption{\label{tab:1}The simulation results for the global polarizaton of
the $\Lambda$ hyperon. We set $\mathscr{P}_{\Lambda}=1/3$, i.e.
the $\Lambda$ hyperons are complete polarized. By analyzing the momentum
distribution of daughter protons in the lab frame, we can determine
the $\Lambda$'s polarization. The results of four methods are presented:
the method by Eqs. (\ref{eq:p-lambda},\ref{eq:spin}) used in the
STAR experiment \cite{Abelev:2007zk} and three by Eqs. (\ref{eq:angular-av},\ref{eq:p-star-1},\ref{eq:angular-av-1})
proposed in this paper. The numbers of events collected are $4\times10^{4}$
at 200 GeV and $2.5\times10^{4}$ at other energies. The results of
method 1-4 are from events in the full rapidity range, while those
of method 5 are in the rapidity range $[-0.5,0.5]$. }

\begin{tabular}{|c|c|c|c|c|c|c|}
\hline 
Energy & method 1 & method 2 & method 3 & method 4 & method 5 & Number of $\Lambda$s\tabularnewline
GeV & Eq. (\ref{eq:p-lambda}) & Eq. (\ref{eq:spin}) & Eq. (\ref{eq:angular-av}) & Eq.(\ref{eq:p-star-1}) & Eq. (\ref{eq:angular-av-1}) & (full rapidity)\tabularnewline
\hline 
\hline 
200 & 0.33581 & 0.335851 & 0.3324 & 0.33014 & 0.308495 & 1304795\tabularnewline
\hline 
180 & 0.330877 & 0.33141 & 0.326565 & 0.329057 & 0.306966 & 927717\tabularnewline
\hline 
140 & 0.338745 & 0.337673 & 0.338942 & 0.335862 & 0.351934 & 892533\tabularnewline
\hline 
120 & 0.333962 & 0.333688 & 0.329696 & 0.334152 & 0.318965 & 995522\tabularnewline
\hline 
100 & 0.336686 & 0.334685 & 0.34669 & 0.34522 & 0.360992 & 971596\tabularnewline
\hline 
62.4 & 0.331964 & 0.33118 & 0.324133 & 0.333466 & 0.353216  & 918787\tabularnewline
\hline 
40 & 0.330536 & 0.330302 & 0.332092 & 0.331782 & 0.323459 & 795837\tabularnewline
\hline 
39 & 0.337252 & 0.337516 & 0.332983 & 0.331683 & 0.312195  & 847367\tabularnewline
\hline 
19.6 & 0.328531 & 0.328434 & 0.339587 & 0.328939 & 0.31276  & 707868\tabularnewline
\hline 
7.7 & 0.341257 & 0.3417 & 0.364069 & 0.34862 & 0.302301  & 434697\tabularnewline
\hline 
\end{tabular}
\end{table}

\section{Summary}

The previous method used in the STAR experiment to measure the global
$\Lambda$ polarization is through the event average of $\sin\left(\phi_{\mathrm{p}}^{*}-\psi_{\mathrm{RP}}\right)$,
where $\phi_{\mathrm{p}}^{*}$ and $\psi_{\mathrm{RP}}$ are the azimuthal
angle of the proton momemtum in $\Lambda$'s rest frame and that of
the reaction plane respectively. We propose several alternative methods
to measure the global $\Lambda$ polarization in the lab frame. Based
on Lorentz transformation for momenta, we can express the gobal polarization
in terms of momenta of protons and $\Lambda$ hyperons in the lab
frame. So the event average can be taken over quantities in the lab
frame. To test how well these methods are for measuring the global
polarization compared to the STAR's method, we use the UrQMD model
to produce an ensemble of $\Lambda$'s momenta and then sample the
angular distribution of protons and pions following the weak decay
formula for $\Lambda$ hyperons. By taking event average over quantities
as functions of momenta of protons and $\Lambda$s in the lab frame
we can determine the global polarization by these methods as well
as by the STAR's. The simulation results show that all these methods
work equivalently well as the STAR method. 

\textit{Acknowledgments.} 

IS and QW are supported in part by the Major State Basic Research
Development Program (MSBRD) in China under the Grant No. 2015CB856902
and 2014CB845402 and by the National Natural Science Foundation of
China (NSFC) under the Grant No. 11535012.  

\bibliographystyle{apsrev}
\bibliography{ref-1}

\begin{thebibliography}{32}
\expandafter\ifx\csname natexlab\endcsname\relax\def\natexlab#1{#1}\fi
\expandafter\ifx\csname bibnamefont\endcsname\relax
  \def\bibnamefont#1{#1}\fi
\expandafter\ifx\csname bibfnamefont\endcsname\relax
  \def\bibfnamefont#1{#1}\fi
\expandafter\ifx\csname citenamefont\endcsname\relax
  \def\citenamefont#1{#1}\fi
\expandafter\ifx\csname url\endcsname\relax
  \def\url#1{\texttt{#1}}\fi
\expandafter\ifx\csname urlprefix\endcsname\relax\def\urlprefix{URL }\fi
\providecommand{\bibinfo}[2]{#2}
\providecommand{\eprint}[2][]{\url{#2}}

\bibitem[{\citenamefont{Barnett}(1935)}]{Barnett:1935}
\bibinfo{author}{\bibfnamefont{S.}~\bibnamefont{Barnett}},
  \bibinfo{journal}{Rev. Mod. Rev.} \textbf{\bibinfo{volume}{7}},
  \bibinfo{pages}{129} (\bibinfo{year}{1935}).

\bibitem[{\citenamefont{Einstein and de~Haas}(1915)}]{dehaas:1915}
\bibinfo{author}{\bibfnamefont{A.}~\bibnamefont{Einstein}} \bibnamefont{and}
  \bibinfo{author}{\bibfnamefont{W.}~\bibnamefont{de~Haas}},
  \bibinfo{journal}{Deutsche Physikalische Gesellschaft, Verhandlungen}
  \textbf{\bibinfo{volume}{17}}, \bibinfo{pages}{152} (\bibinfo{year}{1915}).

\bibitem[{\citenamefont{Liang and Wang}(2005{\natexlab{a}})}]{Liang:2004ph}
\bibinfo{author}{\bibfnamefont{Z.-T.} \bibnamefont{Liang}} \bibnamefont{and}
  \bibinfo{author}{\bibfnamefont{X.-N.} \bibnamefont{Wang}},
  \bibinfo{journal}{Phys. Rev. Lett.} \textbf{\bibinfo{volume}{94}},
  \bibinfo{pages}{102301} (\bibinfo{year}{2005}{\natexlab{a}}),
  \bibinfo{note}{[Erratum: Phys. Rev. Lett.96,039901(2006)]},
  \eprint{nucl-th/0410079}.

\bibitem[{\citenamefont{Liang and Wang}(2005{\natexlab{b}})}]{Liang:2004xn}
\bibinfo{author}{\bibfnamefont{Z.-T.} \bibnamefont{Liang}} \bibnamefont{and}
  \bibinfo{author}{\bibfnamefont{X.-N.} \bibnamefont{Wang}},
  \bibinfo{journal}{Phys. Lett.} \textbf{\bibinfo{volume}{B629}},
  \bibinfo{pages}{20} (\bibinfo{year}{2005}{\natexlab{b}}),
  \eprint{nucl-th/0411101}.

\bibitem[{\citenamefont{Voloshin}(2004)}]{Voloshin:2004ha}
\bibinfo{author}{\bibfnamefont{S.~A.} \bibnamefont{Voloshin}}
  (\bibinfo{year}{2004}), \eprint{nucl-th/0410089}.

\bibitem[{\citenamefont{Betz et~al.}(2007)\citenamefont{Betz, Gyulassy, and
  Torrieri}}]{Betz2007}
\bibinfo{author}{\bibfnamefont{B.}~\bibnamefont{Betz}},
  \bibinfo{author}{\bibfnamefont{M.}~\bibnamefont{Gyulassy}}, \bibnamefont{and}
  \bibinfo{author}{\bibfnamefont{G.}~\bibnamefont{Torrieri}},
  \bibinfo{journal}{Phys. Rev.} \textbf{\bibinfo{volume}{C76}},
  \bibinfo{pages}{044901} (\bibinfo{year}{2007}), \eprint{0708.0035}.

\bibitem[{\citenamefont{Becattini et~al.}(2008)\citenamefont{Becattini,
  Piccinini, and Rizzo}}]{Becattini:2007sr}
\bibinfo{author}{\bibfnamefont{F.}~\bibnamefont{Becattini}},
  \bibinfo{author}{\bibfnamefont{F.}~\bibnamefont{Piccinini}},
  \bibnamefont{and} \bibinfo{author}{\bibfnamefont{J.}~\bibnamefont{Rizzo}},
  \bibinfo{journal}{Phys. Rev.} \textbf{\bibinfo{volume}{C77}},
  \bibinfo{pages}{024906} (\bibinfo{year}{2008}), \eprint{0711.1253}.

\bibitem[{\citenamefont{Gao et~al.}(2008)\citenamefont{Gao, Chen, Deng, Liang,
  Wang, and Wang}}]{Gao2008}
\bibinfo{author}{\bibfnamefont{J.-H.} \bibnamefont{Gao}},
  \bibinfo{author}{\bibfnamefont{S.-W.} \bibnamefont{Chen}},
  \bibinfo{author}{\bibfnamefont{W.-t.} \bibnamefont{Deng}},
  \bibinfo{author}{\bibfnamefont{Z.-T.} \bibnamefont{Liang}},
  \bibinfo{author}{\bibfnamefont{Q.}~\bibnamefont{Wang}}, \bibnamefont{and}
  \bibinfo{author}{\bibfnamefont{X.-N.} \bibnamefont{Wang}},
  \bibinfo{journal}{Phys. Rev.} \textbf{\bibinfo{volume}{C77}},
  \bibinfo{pages}{044902} (\bibinfo{year}{2008}), \eprint{0710.2943}.

\bibitem[{\citenamefont{Chen et~al.}(2009)\citenamefont{Chen, Deng, Gao, and
  Wang}}]{Chen:2008wh}
\bibinfo{author}{\bibfnamefont{S.-w.} \bibnamefont{Chen}},
  \bibinfo{author}{\bibfnamefont{J.}~\bibnamefont{Deng}},
  \bibinfo{author}{\bibfnamefont{J.-h.} \bibnamefont{Gao}}, \bibnamefont{and}
  \bibinfo{author}{\bibfnamefont{Q.}~\bibnamefont{Wang}},
  \bibinfo{journal}{Front. Phys. China} \textbf{\bibinfo{volume}{4}},
  \bibinfo{pages}{509} (\bibinfo{year}{2009}), \eprint{0801.2296}.

\bibitem[{\citenamefont{Becattini and Tinti}(2010)}]{Becattini:2009wh}
\bibinfo{author}{\bibfnamefont{F.}~\bibnamefont{Becattini}} \bibnamefont{and}
  \bibinfo{author}{\bibfnamefont{L.}~\bibnamefont{Tinti}},
  \bibinfo{journal}{Annals Phys.} \textbf{\bibinfo{volume}{325}},
  \bibinfo{pages}{1566} (\bibinfo{year}{2010}), \eprint{0911.0864}.

\bibitem[{\citenamefont{Becattini}(2012)}]{Becattini:2012tc}
\bibinfo{author}{\bibfnamefont{F.}~\bibnamefont{Becattini}},
  \bibinfo{journal}{Phys. Rev. Lett.} \textbf{\bibinfo{volume}{108}},
  \bibinfo{pages}{244502} (\bibinfo{year}{2012}), \eprint{1201.5278}.

\bibitem[{\citenamefont{Becattini et~al.}(2013)\citenamefont{Becattini,
  Chandra, Del~Zanna, and Grossi}}]{Becattini:2013fla}
\bibinfo{author}{\bibfnamefont{F.}~\bibnamefont{Becattini}},
  \bibinfo{author}{\bibfnamefont{V.}~\bibnamefont{Chandra}},
  \bibinfo{author}{\bibfnamefont{L.}~\bibnamefont{Del~Zanna}},
  \bibnamefont{and} \bibinfo{author}{\bibfnamefont{E.}~\bibnamefont{Grossi}},
  \bibinfo{journal}{Annals Phys.} \textbf{\bibinfo{volume}{338}},
  \bibinfo{pages}{32} (\bibinfo{year}{2013}), \eprint{1303.3431}.

\bibitem[{\citenamefont{Becattini and Grossi}(2015)}]{Becattini:2015nva}
\bibinfo{author}{\bibfnamefont{F.}~\bibnamefont{Becattini}} \bibnamefont{and}
  \bibinfo{author}{\bibfnamefont{E.}~\bibnamefont{Grossi}},
  \bibinfo{journal}{Phys. Rev.} \textbf{\bibinfo{volume}{D92}},
  \bibinfo{pages}{045037} (\bibinfo{year}{2015}), \eprint{1505.07760}.

\bibitem[{\citenamefont{Gao et~al.}(2012)\citenamefont{Gao, Liang, Pu, Wang,
  and Wang}}]{Gao:2012ix}
\bibinfo{author}{\bibfnamefont{J.-H.} \bibnamefont{Gao}},
  \bibinfo{author}{\bibfnamefont{Z.-T.} \bibnamefont{Liang}},
  \bibinfo{author}{\bibfnamefont{S.}~\bibnamefont{Pu}},
  \bibinfo{author}{\bibfnamefont{Q.}~\bibnamefont{Wang}}, \bibnamefont{and}
  \bibinfo{author}{\bibfnamefont{X.-N.} \bibnamefont{Wang}},
  \bibinfo{journal}{Phys.Rev.Lett.} \textbf{\bibinfo{volume}{109}},
  \bibinfo{pages}{232301} (\bibinfo{year}{2012}), \eprint{1203.0725}.

\bibitem[{\citenamefont{Chen et~al.}(2013)\citenamefont{Chen, Pu, Wang, and
  Wang}}]{Chen:2012ca}
\bibinfo{author}{\bibfnamefont{J.-W.} \bibnamefont{Chen}},
  \bibinfo{author}{\bibfnamefont{S.}~\bibnamefont{Pu}},
  \bibinfo{author}{\bibfnamefont{Q.}~\bibnamefont{Wang}}, \bibnamefont{and}
  \bibinfo{author}{\bibfnamefont{X.-N.} \bibnamefont{Wang}},
  \bibinfo{journal}{Phys. Rev. Lett.} \textbf{\bibinfo{volume}{110}},
  \bibinfo{pages}{262301} (\bibinfo{year}{2013}), \eprint{1210.8312}.

\bibitem[{\citenamefont{Fang et~al.}(2016)\citenamefont{Fang, Pang, Wang, and
  Wang}}]{Fang:2016vpj}
\bibinfo{author}{\bibfnamefont{R.-h.} \bibnamefont{Fang}},
  \bibinfo{author}{\bibfnamefont{L.-g.} \bibnamefont{Pang}},
  \bibinfo{author}{\bibfnamefont{Q.}~\bibnamefont{Wang}}, \bibnamefont{and}
  \bibinfo{author}{\bibfnamefont{X.-n.} \bibnamefont{Wang}},
  \bibinfo{journal}{Phys. Rev.} \textbf{\bibinfo{volume}{C94}},
  \bibinfo{pages}{024904} (\bibinfo{year}{2016}), \eprint{1604.04036}.

\bibitem[{\citenamefont{Fang et~al.}(2017)\citenamefont{Fang, Pang, Wang, and
  Wang}}]{Fang:2016uds}
\bibinfo{author}{\bibfnamefont{R.-h.} \bibnamefont{Fang}},
  \bibinfo{author}{\bibfnamefont{J.-y.} \bibnamefont{Pang}},
  \bibinfo{author}{\bibfnamefont{Q.}~\bibnamefont{Wang}}, \bibnamefont{and}
  \bibinfo{author}{\bibfnamefont{X.-n.} \bibnamefont{Wang}},
  \bibinfo{journal}{Phys. Rev.} \textbf{\bibinfo{volume}{D95}},
  \bibinfo{pages}{014032} (\bibinfo{year}{2017}), \eprint{1611.04670}.

\bibitem[{\citenamefont{Wang}(2017)}]{Wang:2017jpl}
\bibinfo{author}{\bibfnamefont{Q.}~\bibnamefont{Wang}}, in
  \emph{\bibinfo{booktitle}{{26th International Conference on Ultrarelativistic
  Nucleus-Nucleus Collisions (Quark Matter 2017) Chicago,Illinois, USA,
  February 6-11, 2017}}} (\bibinfo{year}{2017}), \eprint{1704.04022},
  \urlprefix\url{http://inspirehep.net/record/1591567/files/arXiv:1704.04022.pdf}.

\bibitem[{\citenamefont{Overseth and Roth}(1967)}]{Overseth:1967zz}
\bibinfo{author}{\bibfnamefont{O.~E.} \bibnamefont{Overseth}} \bibnamefont{and}
  \bibinfo{author}{\bibfnamefont{R.~F.} \bibnamefont{Roth}},
  \bibinfo{journal}{Phys. Rev. Lett.} \textbf{\bibinfo{volume}{19}},
  \bibinfo{pages}{391} (\bibinfo{year}{1967}).

\bibitem[{\citenamefont{Adamczyk et~al.}(2017)}]{STAR:2017ckg}
\bibinfo{author}{\bibfnamefont{L.}~\bibnamefont{Adamczyk}} \bibnamefont{et~al.}
  (\bibinfo{collaboration}{STAR}), \bibinfo{journal}{Nature}
  \textbf{\bibinfo{volume}{548}}, \bibinfo{pages}{62} (\bibinfo{year}{2017}),
  \eprint{1701.06657}.

\bibitem[{\citenamefont{Abelev et~al.}(2007)}]{Abelev:2007zk}
\bibinfo{author}{\bibfnamefont{B.~I.} \bibnamefont{Abelev}}
  \bibnamefont{et~al.} (\bibinfo{collaboration}{STAR}), \bibinfo{journal}{Phys.
  Rev.} \textbf{\bibinfo{volume}{C76}}, \bibinfo{pages}{024915}
  (\bibinfo{year}{2007}), \eprint{0705.1691}.

\bibitem[{\citenamefont{Csernai et~al.}(2013)\citenamefont{Csernai, Magas, and
  Wang}}]{Csernai:2013bqa}
\bibinfo{author}{\bibfnamefont{L.~P.} \bibnamefont{Csernai}},
  \bibinfo{author}{\bibfnamefont{V.~K.} \bibnamefont{Magas}}, \bibnamefont{and}
  \bibinfo{author}{\bibfnamefont{D.~J.} \bibnamefont{Wang}},
  \bibinfo{journal}{Phys. Rev.} \textbf{\bibinfo{volume}{C87}},
  \bibinfo{pages}{034906} (\bibinfo{year}{2013}), \eprint{1302.5310}.

\bibitem[{\citenamefont{Csernai et~al.}(2014)\citenamefont{Csernai, Wang,
  Bleicher, and Stoecker}}]{Csernai:2014ywa}
\bibinfo{author}{\bibfnamefont{L.~P.} \bibnamefont{Csernai}},
  \bibinfo{author}{\bibfnamefont{D.~J.} \bibnamefont{Wang}},
  \bibinfo{author}{\bibfnamefont{M.}~\bibnamefont{Bleicher}}, \bibnamefont{and}
  \bibinfo{author}{\bibfnamefont{H.}~\bibnamefont{Stoecker}},
  \bibinfo{journal}{Phys. Rev.} \textbf{\bibinfo{volume}{C90}},
  \bibinfo{pages}{021904} (\bibinfo{year}{2014}).

\bibitem[{\citenamefont{Pang et~al.}(2016)\citenamefont{Pang, Petersen, Wang,
  and Wang}}]{Pang:2016igs}
\bibinfo{author}{\bibfnamefont{L.-G.} \bibnamefont{Pang}},
  \bibinfo{author}{\bibfnamefont{H.}~\bibnamefont{Petersen}},
  \bibinfo{author}{\bibfnamefont{Q.}~\bibnamefont{Wang}}, \bibnamefont{and}
  \bibinfo{author}{\bibfnamefont{X.-N.} \bibnamefont{Wang}},
  \bibinfo{journal}{Phys. Rev. Lett.} \textbf{\bibinfo{volume}{117}},
  \bibinfo{pages}{192301} (\bibinfo{year}{2016}), \eprint{1605.04024}.

\bibitem[{\citenamefont{Jiang et~al.}(2016)\citenamefont{Jiang, Lin, and
  Liao}}]{Jiang:2016woz}
\bibinfo{author}{\bibfnamefont{Y.}~\bibnamefont{Jiang}},
  \bibinfo{author}{\bibfnamefont{Z.-W.} \bibnamefont{Lin}}, \bibnamefont{and}
  \bibinfo{author}{\bibfnamefont{J.}~\bibnamefont{Liao}},
  \bibinfo{journal}{Phys. Rev.} \textbf{\bibinfo{volume}{C94}},
  \bibinfo{pages}{044910} (\bibinfo{year}{2016}), \eprint{1602.06580}.

\bibitem[{\citenamefont{Deng and Huang}(2016)}]{Deng:2016gyh}
\bibinfo{author}{\bibfnamefont{W.-T.} \bibnamefont{Deng}} \bibnamefont{and}
  \bibinfo{author}{\bibfnamefont{X.-G.} \bibnamefont{Huang}},
  \bibinfo{journal}{Phys. Rev.} \textbf{\bibinfo{volume}{C93}},
  \bibinfo{pages}{064907} (\bibinfo{year}{2016}), \eprint{1603.06117}.

\bibitem[{\citenamefont{Karpenko and Becattini}(2017)}]{Karpenko:2016jyx}
\bibinfo{author}{\bibfnamefont{I.}~\bibnamefont{Karpenko}} \bibnamefont{and}
  \bibinfo{author}{\bibfnamefont{F.}~\bibnamefont{Becattini}},
  \bibinfo{journal}{Eur. Phys. J.} \textbf{\bibinfo{volume}{C77}},
  \bibinfo{pages}{213} (\bibinfo{year}{2017}), \eprint{1610.04717}.

\bibitem[{\citenamefont{Xie et~al.}(2017)\citenamefont{Xie, Wang, and
  Csernai}}]{Xie:2017upb}
\bibinfo{author}{\bibfnamefont{Y.}~\bibnamefont{Xie}},
  \bibinfo{author}{\bibfnamefont{D.}~\bibnamefont{Wang}}, \bibnamefont{and}
  \bibinfo{author}{\bibfnamefont{L.~P.} \bibnamefont{Csernai}},
  \bibinfo{journal}{Phys. Rev.} \textbf{\bibinfo{volume}{C95}},
  \bibinfo{pages}{031901} (\bibinfo{year}{2017}), \eprint{1703.03770}.

\bibitem[{\citenamefont{Li et~al.}(2017)\citenamefont{Li, Pang, Wang, and
  Xia}}]{Li:2017slc}
\bibinfo{author}{\bibfnamefont{H.}~\bibnamefont{Li}},
  \bibinfo{author}{\bibfnamefont{L.-G.} \bibnamefont{Pang}},
  \bibinfo{author}{\bibfnamefont{Q.}~\bibnamefont{Wang}}, \bibnamefont{and}
  \bibinfo{author}{\bibfnamefont{X.-L.} \bibnamefont{Xia}}
  (\bibinfo{year}{2017}), \eprint{1704.01507}.

\bibitem[{\citenamefont{Sun and Ko}(2017)}]{Sun:2017xhx}
\bibinfo{author}{\bibfnamefont{Y.}~\bibnamefont{Sun}} \bibnamefont{and}
  \bibinfo{author}{\bibfnamefont{C.~M.} \bibnamefont{Ko}},
  \bibinfo{journal}{Phys. Rev.} \textbf{\bibinfo{volume}{C96}},
  \bibinfo{pages}{024906} (\bibinfo{year}{2017}), \eprint{1706.09467}.

\bibitem[{\citenamefont{Bass et~al.}(1998)}]{Bass:1998ca}
\bibinfo{author}{\bibfnamefont{S.~A.} \bibnamefont{Bass}} \bibnamefont{et~al.},
  \bibinfo{journal}{Prog. Part. Nucl. Phys.} \textbf{\bibinfo{volume}{41}},
  \bibinfo{pages}{255} (\bibinfo{year}{1998}), \bibinfo{note}{[Prog. Part.
  Nucl. Phys.41,225(1998)]}, \eprint{nucl-th/9803035}.

\bibitem[{\citenamefont{Petersen et~al.}(2008)\citenamefont{Petersen,
  Steinheimer, Burau, Bleicher, and Stocker}}]{Petersen:2008dd}
\bibinfo{author}{\bibfnamefont{H.}~\bibnamefont{Petersen}},
  \bibinfo{author}{\bibfnamefont{J.}~\bibnamefont{Steinheimer}},
  \bibinfo{author}{\bibfnamefont{G.}~\bibnamefont{Burau}},
  \bibinfo{author}{\bibfnamefont{M.}~\bibnamefont{Bleicher}}, \bibnamefont{and}
  \bibinfo{author}{\bibfnamefont{H.}~\bibnamefont{Stocker}},
  \bibinfo{journal}{Phys. Rev.} \textbf{\bibinfo{volume}{C78}},
  \bibinfo{pages}{044901} (\bibinfo{year}{2008}), \eprint{0806.1695}.

\end{thebibliography}

\end{document}